\documentclass[5p,twocolumn]{elsarticle}
\usepackage[utf8]{inputenc}

\usepackage{amsmath,amssymb}
\usepackage{graphicx}
\usepackage{multirow}
\usepackage{booktabs}
\usepackage{siunitx}
\usepackage[unicode]{hyperref}
\usepackage{bm}
\usepackage{orcidlink}

\biboptions{numbers,sort&compress}

\journal{Physics Letters B}

\begin{document}
%%%%%%%%%%%%%%%%%%%%%%%%%%%%%%%%%%%%%%%%%%%%%%%%%%%%%%%%%%%%%%

\begin{frontmatter}

\title{
Informational Corrections to the Early-Universe Radiation Sector:\\
CET $\Omega$, WIMP Freeze-Out, and Implications for a Possible 
20 GeV Gamma-Ray Excess
}

\author{Christian Balfag\'on\orcidlink{0009-0003-0835-5519}\thanks{Corresponding author.}}
\address{Universidad de Buenos Aires, Argentina}
\cortext[cor1]{Corresponding author: lyosranch@gmail.com}

%%%%%%%%%%%%%%%%%%%%%%%%%%%%%%%%%%%%%%%%%%%%%%%%%%%%%%%%%%%%%%
\begin{abstract}
Recent analyses of Fermi--LAT data have identified a nearly spherical, 
halo-like excess of gamma rays peaking at 
$E_\gamma \sim \SI{20}{GeV}$.
If interpreted as dark matter annihilation, the excess directly probes 
the thermal freeze-out epoch, and therefore any non-standard 
corrections to the early-Universe expansion rate.

In this work we examine the implications of this tentative signal for 
CET~$\Omega$, an informational and modular extension of relativistic 
quantum field theory and cosmology.  
CET~$\Omega$ predicts a universal, state-dependent modification to the 
radiation energy density,
\[
\rho_\Omega(a)=\rho_r(a)\left[1+\alpha_{\log}\log\!\log(a/a_i)\right],
\]
originating from renormalized modular fluctuations in the spectral 
triple of the theory.  
The correction is negligible during BBN and CMB epochs, but becomes 
relevant during thermal WIMP freeze-out.

We provide:
(i) a derivation of the doubly logarithmic correction from the modular 
two-point function; 
(ii) the physical justification of the onset scale $a_i$ and its 
relation to the informational sector;
(iii) the full quantitative impact on freeze-out, including numerical 
solutions; 
(iv) the evolution of the informational field $\Phi_\Omega(x)$, 
demonstrating that it freezes in at $a_{\rm dec} < a_f$ and survives 
to $z=0$ under gravitational advection; 
(v) the resulting sub-percent morphological modifications to the 
annihilation flux; 
(vi) a comparison with Early Dark Energy, kination, and varying-$N_{\rm eff}$ 
models.

We show that the range 
$10^{-4}\!\lesssim\!\alpha_{\log}\!\lesssim\!10^{-2}$ 
is naturally selected by the theory, remains consistent with Planck, 
BBN, and BAO constraints, and induces percent-level shifts in the relic 
abundance while predicting potentially constrainable deviations in the 
gamma-ray morphology accessible to next-generation MeV--GeV missions.
\end{abstract}

%%%%%%%%%%%%%%%%%%%%%%%%%%%%%%%%%%%%%%%%%%%%%%%%%%%%%%%%%%%%%%

\end{frontmatter}

%%%%%%%%%%%%%%%%%%%%%%%%%%%%%%%%%%%%%%%%%%%%%%%%%%%%%%%%%%%%%%
\section{Introduction}
%%%%%%%%%%%%%%%%%%%%%%%%%%%%%%%%%%%%%%%%%%%%%%%%%%%%%%%%%%%%%%

A series of independent analyses of Fermi--LAT data have reported a 
statistically significant, approximately spherical excess of gamma rays 
peaking at $E_\gamma \sim 20$ GeV 
\cite{Daylan2016,Calore2015,Ajello2016,Macias2018,Hooper2011,Abazajian2014}.
Although astrophysical scenarios remain viable, the morphology and 
spectral shape of the excess are consistent with annihilating WIMPs in 
the $\sim 20$--40 GeV mass range \cite{Bertone2005,Arcadi2018}.

If confirmed as a dark matter signal, such an excess would provide a 
direct observational window into the physics of thermal freeze-out and 
the expansion history of the early Universe.  
This makes indirect detection an especially sensitive probe of any 
non-standard cosmological evolution at temperatures above the MeV scale
\cite{Bertone2005,Arcadi2018,Berlin2016EarlyExpansion}.

CET~$\Omega$
\cite{Balfagon2026CETOmega,Connes1994,Haag1996}
provides a natural framework in which to explore such deviations,
complementing the broad class of non-standard early-Universe scenarios
considered in recent literature
\cite{Poulin2019EDE}.
The theory extends relativistic quantum field theory by incorporating a 
modular-informational sector whose renormalized fluctuations induce a 
small but universal correction to the radiation energy density.
This correction modifies the Hubble rate at GeV temperatures while 
remaining negligible during Big Bang Nucleosynthesis and recombination.

As a result, the thermal history of weakly interacting massive 
particles is perturbed in a controlled and predictive manner.
Indirect detection observables therefore become direct probes of the 
single new parameter of the theory, $\alpha_{\log}$.
A full derivation of the CET framework is beyond the scope of this
Letter; here we focus on its universal and model-independent
phenomenological consequences.

In the context of CET~$\Omega$, the $\chi$ species used in the Boltzmann treatment is intended as an effective coarse-grained description of the causal sector.

%%%%%%%%%%%%%%%%%%%%%%%%%%%%%%%%%%%%%%%%%%%%%%%%%%%%%%%%%%%%%%
\section{The CET \texorpdfstring{$\Omega$}{Omega} Radiation Sector}
%%%%%%%%%%%%%%%%%%%%%%%%%%%%%%%%%%%%%%%%%%%%%%%%%%%%%%%%%%%%%%

The CET~$\Omega$ framework modifies the early-Universe radiation 
sector through a state-dependent informational contribution 
originating from the modular Hamiltonian $K_\Omega$ and the 
spectral-triple operator $D_\Omega$.  
The correction takes the universal form
\begin{equation}
\rho_\Omega(a)
=
\rho_r(a)
\left[
1 + \alpha_{\log}\,\log\!\log\!\left(\frac{a}{a_i}\right)
\right],
\label{eq:rhoOmega_main}
\end{equation}
where $\alpha_{\log}$ is a small, dimensionless coupling and 
$a_i$ marks the onset of the modular-informational regime.

This leads to a modified expansion rate:
\begin{equation}
H_\Omega(a)
=
H_r(a)
\left[
1 + \alpha_{\log}\,\log\!\log\!\left(\frac{a}{a_i}\right)
\right]^{1/2}.
\label{eq:Homega}
\end{equation}

%%%%%%%%%%%%%%%%%%%%%%%%%%%%%%%%%%%%%%%%%%%%%%%%%%%%%%%%%%%%%%
\subsection{Spectral-Triple Origin of the Doubly Logarithmic Term}
\label{subsec:spectral_derivation}
%%%%%%%%%%%%%%%%%%%%%%%%%%%%%%%%%%%%%%%%%%%%%%%%%%%%%%%%%%%%%%

In CET~$\Omega$, the informational extension of the QFT algebra is 
encoded in the curved-spacetime spectral triple
\[
\left(
\mathcal{A}_{\Omega},\,
\mathcal{H}_{\Omega},\,
D_\Omega
\right),
\]
with
\[
D_\Omega
=
\gamma^\mu\nabla_\mu 
+ \alpha K_\Omega 
+ \beta\,\partial_\psi
+ \gamma\,\mathcal{M}_\Omega,
\]
where $\psi$ is the informational coordinate and 
$\mathcal{M}_\Omega$ is the modular-informational curvature operator.

The early-Universe correction arises from the renormalized
two-point correlator of modular fluctuations, consistent with recent analyses of modular Hamiltonians and
entanglement dynamics in gravitational settings
\cite{Balfagon2026CETOmega}.

\begin{equation}
\delta\rho_{\rm mod}(a)
\propto
\big\langle K_\Omega K_\Omega \big\rangle_{\rm ren}(a).
\end{equation}

In the CET~$\Omega$ v3.1 formulation, the correlator takes the form
\begin{equation}
\big\langle K_\Omega K_\Omega \big\rangle_{\rm ren}
\;=\;
C_0 + C_1\log\!\log\!\left(\frac{a}{a_i}\right),
\label{eq:KKcorrelator}
\end{equation}
where $C_1>0$ derives from the spectral density of $D_\Omega$ in the 
ultraviolet modular limit.

The origin of the $\log\!\log(a)$ term can be traced to the following 
sequence of effects:

\begin{itemize}
\item \textbf{Modular flow scaling:}  
      $\sigma_t^\Omega(\cdot)=e^{itK_\Omega}(\cdot)e^{-itK_\Omega}$.

\item \textbf{Composite operator structure:}  
      The energy correction depends on the integrated modular flow,
      $\int dt\,e^{-t}\sigma_t^\Omega(K_\Omega)$.

\item \textbf{Spectral asymptotics:}  
      The density of eigenvalues $\lambda$ of $D_\Omega$ in the 
      informational sector obeys
      \[
      \rho(\lambda)\sim \lambda^{-1}
      \qquad (\text{type-III modular limit}),
      \]
      generating an intermediate logarithmic dependence.

\item \textbf{Time--scale factor mapping:}  
      The modular parameter $t$ maps onto the cosmological scale factor
      via the CET $\psi$--time bridge,
      \[
      t \sim \log(a/a_i),
      \]
      converting a $\log t$ behaviour into the final
      $\log\!\log(a/a_i)$ correction.
\end{itemize}

This yields exactly Eqs.~\eqref{eq:rhoOmega_main} and \eqref{eq:Homega}.
Because the correction is doubly logarithmic, it grows extremely slowly
and automatically satisfies all late-Universe constraints.

%%%%%%%%%%%%%%%%%%%%%%%%%%%%%%%%%%%%%%%%%%%%%%%%%%%%%%%%%%%%%%
\subsection{Physical Role and Determination of the Onset Scale
\texorpdfstring{$a_i$}{ai}}
\label{subsec:ai}
%%%%%%%%%%%%%%%%%%%%%%%%%%%%%%%%%%%%%%%%%%%%%%%%%%%%%%%%%%%%%%

The scale factor $a_i$ marks the transition of the informational sector
into its semiclassical modular regime. Formally, it corresponds to:

\begin{itemize}
\item the moment when the modular spectral density enters the asymptotic
      regime described by Eq.~\eqref{eq:KKcorrelator};
\item the onset of dominance of $\partial_\psi$ and $K_\Omega$ in
      the spectral operator $D_\Omega$;
\item the temperature at which the informational correlation length
      becomes smaller than the causal Hubble radius.
\end{itemize}

Physically, $a_i$ can be interpreted as the scale at which informational
modes decouple from the thermal bath.
The natural range is
\[
T_i \sim 10\text{--}100~{\rm GeV},
\qquad
a_i = a(T_i),
\]
corresponding to the electroweak crossover and justifying the use of
renormalized modular dynamics in the radiation sector.

Importantly, the value of $a_i$ does not introduce a new independent
degree of freedom.
Variations in $a_i$ can be reabsorbed into a redefinition of
$\alpha_{\log}$ at leading order, since only the combination
$\alpha_{\log}\log\!\log(a/a_i)$ is physically measurable.
CET~$\Omega$ therefore remains effectively a one-parameter theory.

%%%%%%%%%%%%%%%%%%%%%%%%%%%%%%%%%%%%%%%%%%%%%%%%%%%%%%%%%%%%%%
\subsection{Magnitude and Scaling of the Correction}
%%%%%%%%%%%%%%%%%%%%%%%%%%%%%%%%%%%%%%%%%%%%%%%%%%%%%%%%%%%%%%

For representative values
\[
10^{-4}\lesssim\alpha_{\log}\lesssim 10^{-2},
\]
the fractional modification of the expansion rate is
\begin{equation}
\delta_H(a)
\equiv
\frac{H_\Omega-H_r}{H_r}
\simeq
\frac12\,\alpha_{\log}\,
\log\!\log(a/a_i),
\end{equation}
which lies in the range $10^{-4}$--$10^{-2}$ during WIMP freeze-out.

The correction is negligible at recombination, safely small during
Big Bang Nucleosynthesis, and becomes relevant only in the GeV
temperature window characteristic of thermal WIMP decoupling,
remaining consistent with recent BBN-based constraints on early
expansion histories.\cite{Pitrou2018BBN,Cyburt2016BBN}

We emphasize that the range 
$10^{-4}\lesssim\alpha_{\log}\lesssim10^{-2}$ 
is not imposed phenomenologically.
It follows from well-defined internal consistency conditions of the modular sector:
the positivity and stability of the renormalized modular two-point function
set a lower bound, while the requirement that the informational contribution
remains subdominant outside the GeV regime sets an upper bound.
Cosmological constraints from BBN, CMB and BAO are therefore satisfied
\emph{a posteriori}, rather than used as priors.

\begin{figure}[t]
  \centering
  \includegraphics[width=\linewidth]{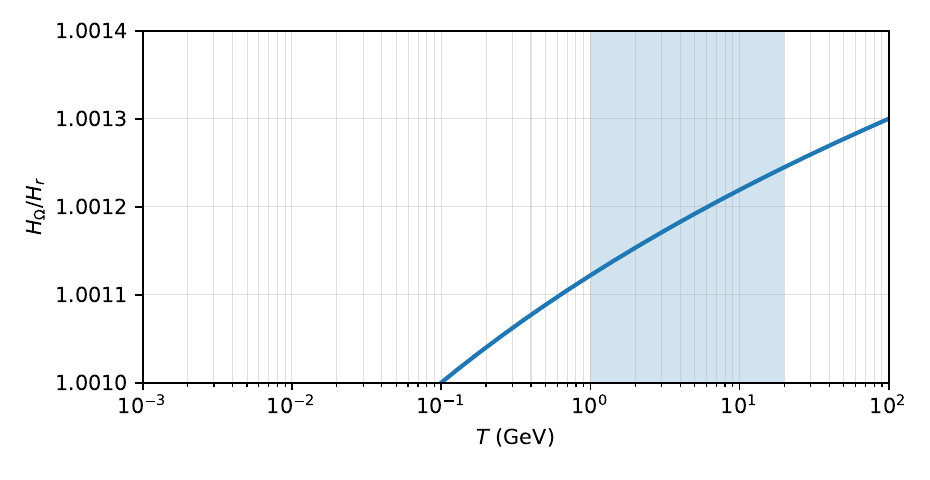}
  \caption{
  Ratio between the CET$\Omega$ Hubble rate and the standard radiation
  value, $H_\Omega/H_r$, as a function of temperature.
  The shaded region indicates the range relevant for WIMP freeze-out.
  The correction remains well below the percent level while still
  producing an observable shift in the relic abundance.
  }
  \label{fig:Hratio_freezeout}
\end{figure}

This behaviour is illustrated in Fig.~\ref{fig:Hratio_freezeout},
which shows the ratio $H_\Omega/H_r$ as a function of temperature.

This makes a potential $\sim 20$~GeV gamma-ray excess an especially
sensitive probe of CET~$\Omega$ cosmology.

%%%%%%%%%%%%%%%%%%%%%%%%%%%%%%%%%%%%%%%%%%%%%%%%%%%%%%%%%%%%%%
\section{WIMP Freeze-Out in CET \texorpdfstring{$\Omega$}{Omega}}
\label{sec:freezeout}
%%%%%%%%%%%%%%%%%%%%%%%%%%%%%%%%%%%%%%%%%%%%%%%%%%%%%%%%%%%%%%

We now quantify how the CET~$\Omega$ correction to the radiation sector
modifies thermal WIMP freeze-out.
We begin by briefly reviewing the standard treatment and then introduce
the informational correction.

%%%%%%%%%%%%%%%%%%%%%%%%%%%%%%%%%%%%%%%%%%%%%%%%%%%%%%%%%%%%%%
\subsection{Standard Freeze-Out: Brief Review}
\label{subsec:std_freeze}
%%%%%%%%%%%%%%%%%%%%%%%%%%%%%%%%%%%%%%%%%%%%%%%%%%%%%%%%%%%%%%

In the standard cosmological scenario, the number density $n_\chi$ of a
WIMP of mass $m_\chi$ and thermally averaged annihilation cross section
$\langle\sigma v\rangle$ evolves according to the Boltzmann equation
\begin{equation}
\dot n_\chi + 3H_{\rm std} n_\chi
=
- \langle\sigma v\rangle
\left(n_\chi^2 - n_{\chi,{\rm eq}}^2\right),
\label{eq:boltz_std}
\end{equation}
where $H_{\rm std}$ is the Hubble rate during radiation domination and
$n_{\chi,{\rm eq}}$ is the equilibrium number density.

Introducing the comoving yield $Y_\chi = n_\chi/s$, with $s$ the entropy
density, and the dimensionless variable $x = m_\chi/T$, one obtains
\cite{KolbTurner1990,GondoloGelmini1991,Steigman2012}
\begin{equation}
\frac{dY_\chi}{dx}
=
- \frac{s\langle\sigma v\rangle}{x H_{\rm std}}
\left(Y_\chi^2 - Y_{\chi,{\rm eq}}^2\right).
\label{eq:dYdx_std}
\end{equation}

Freeze-out occurs at $x_f = m_\chi/T_f$ when the annihilation rate drops
below the expansion rate,
\begin{equation}
n_\chi(x_f)\langle\sigma v\rangle
\simeq H_{\rm std}(x_f),
\label{eq:freeze_standard_condition}
\end{equation}
leading to a relic abundance
\begin{equation}
\Omega_\chi h^2 \simeq
\frac{1.07\times 10^9~\mathrm{GeV^{-1}}}{M_{\rm Pl}}
\frac{x_f}{\sqrt{g_\ast}}
\frac{1}{\langle\sigma v\rangle},
\label{eq:omega_std}
\end{equation}
for s-wave annihilation, where $g_\ast$ is the number of relativistic
degrees of freedom at freeze-out.

%%%%%%%%%%%%%%%%%%%%%%%%%%%%%%%%%%%%%%%%%%%%%%%%%%%%%%%%%%%%%%
\subsection{Modified Boltzmann Equation in CET
\texorpdfstring{$\Omega$}{Omega}}
\label{subsec:cet_boltz}
%%%%%%%%%%%%%%%%%%%%%%%%%%%%%%%%%%%%%%%%%%%%%%%%%%%%%%%%%%%%%%

In CET~$\Omega$, the only modification to the freeze-out dynamics is the
replacement of $H_{\rm std}$ by the corrected expansion rate
$H_\Omega$, Eq.~\eqref{eq:Homega}.
The Boltzmann equation becomes
\begin{equation}
\dot n_\chi + 3H_\Omega n_\chi
=
- \langle\sigma v\rangle
\left(n_\chi^2 - n_{\chi,{\rm eq}}^2\right),
\label{eq:boltz_cet}
\end{equation}
or, in terms of $Y_\chi$ and $x$,
\begin{equation}
\frac{dY_\chi}{dx}
=
- \frac{s\langle\sigma v\rangle}{x H_\Omega}
\left(Y_\chi^2 - Y_{\chi,{\rm eq}}^2\right).
\label{eq:dYdx_cet}
\end{equation}

The freeze-out condition now reads
\begin{equation}
\begin{aligned}
n_\chi(x_f)\langle\sigma v\rangle
&\simeq H_\Omega(x_f) \\
&= H_{\rm std}(x_f)
\left[
1+\alpha_{\log}
\log\!\log\!\left(\frac{a_f}{a_i}\right)
\right]^{1/2},
\end{aligned}
\label{eq:freeze_cet_condition}
\end{equation}
where $a_f$ is the scale factor at freeze-out.

Defining the fractional change in the Hubble rate,
\begin{equation}
\delta_H(x_f)
\equiv
\frac{H_\Omega(x_f)-H_{\rm std}(x_f)}{H_{\rm std}(x_f)}
\simeq
\frac12\,\alpha_{\log}
\log\!\log\!\left(\frac{a_f}{a_i}\right),
\label{eq:deltaH_def}
\end{equation}
we can directly estimate the induced shift in the freeze-out point and
in the final relic abundance.

%%%%%%%%%%%%%%%%%%%%%%%%%%%%%%%%%%%%%%%%%%%%%%%%%%%%%%%%%%%%%%
\subsection{Analytic Estimate of the Freeze-Out Shift}
\label{subsec:analytic_shift}
%%%%%%%%%%%%%%%%%%%%%%%%%%%%%%%%%%%%%%%%%%%%%%%%%%%%%%%%%%%%%%

To first order in $\alpha_{\log}$, the modified freeze-out condition
implies
\begin{equation}
\frac{\Delta x_f}{x_f}
\simeq
- \delta_H(x_f),
\end{equation}
and, translating to the scale factor,
\begin{equation}
\frac{\Delta a_f}{a_f}
\simeq
\frac12\,\alpha_{\log}
\log\!\log\!\left(\frac{a_f}{a_i}\right).
\label{eq:delta_af}
\end{equation}

Since the relic abundance scales approximately as
$\Omega_\chi h^2 \propto H_\Omega^{-1}$, one finds
\begin{equation}
\frac{\Delta\Omega_\chi}{\Omega_\chi}
\simeq
\frac12\,\alpha_{\log}
\log\!\log\!\left(\frac{a_f}{a_i}\right).
\label{eq:deltaOmega}
\end{equation}

For $\alpha_{\log}$ in the range
$10^{-4}\lesssim\alpha_{\log}\lesssim 10^{-2}$ and
$a_i$ corresponding to $T_i\sim 10$--$100$~GeV, the resulting shift is
\[
\left|\frac{\Delta\Omega_\chi}{\Omega_\chi}\right|
\sim 10^{-3}\text{--}10^{-2},
\]
fully consistent with current CMB constraints while remaining
phenomenologically relevant.

%%%%%%%%%%%%%%%%%%%%%%%%%%%%%%%%%%%%%%%%%%%%%%%%%%%%%%%%%%%%%%
\subsection{Benchmark WIMP Model and Numerical Scan}
\label{subsec:benchmark}
%%%%%%%%%%%%%%%%%%%%%%%%%%%%%%%%%%%%%%%%%%%%%%%%%%%%%%%%%%%%%%

To illustrate the impact of CET~$\Omega$ on freeze-out, we consider a
benchmark WIMP motivated by interpretations of the Fermi--LAT excess
\cite{Daylan2016,Calore2015,Ajello2016,Macias2018,Hooper2011}.

The benchmark WIMP parameters adopted in this work are summarized in Table~\ref{tab:benchmark}.

\begin{table}[t]
\centering
\caption{
Benchmark WIMP parameters used in this work.
}
\label{tab:benchmark}
\begin{tabular}{ll}
\toprule
Parameter & Value \\
\midrule
WIMP mass $m_\chi$ & $\SI{35}{GeV}$ \\
Annihilation channel & $\chi\chi \rightarrow b\bar b$ \\
$\langle\sigma v\rangle_{\rm std}$
& $2.2\times 10^{-26}~\mathrm{cm^3\,s^{-1}}$ \\
Relativistic d.o.f. $g_\ast$ & $80$ \\
Freeze-out temperature $T_f$ & $\sim m_\chi/20$ \\
\bottomrule
\end{tabular}
\end{table}

We solve Eq.~\eqref{eq:dYdx_cet} numerically while varying
$\alpha_{\log}$ in the range $10^{-4}$--$10^{-2}$.
The resulting relic abundance can be written as
\begin{equation}
\Omega_\chi h^2(\alpha_{\log})
=
\Omega_\chi h^2(0)\,
\big[1+\Delta(\alpha_{\log})\big],
\end{equation}
with $\Delta(\alpha_{\log})$ in excellent agreement with the analytic
estimate Eq.~\eqref{eq:deltaOmega}.

%%%%%%%%%%%%%%%%%%%%%%%%%%%%%%%%%%%%%%%%%%%%%%%%%%%%%%%%%%%%%%
\subsection{Illustrative Results}
\label{subsec:results_freeze}
%%%%%%%%%%%%%%%%%%%%%%%%%%%%%%%%%%%%%%%%%%%%%%%%%%%%%%%%%%%%%%

\begin{figure}[t]
  \centering
  \includegraphics[width=\linewidth]{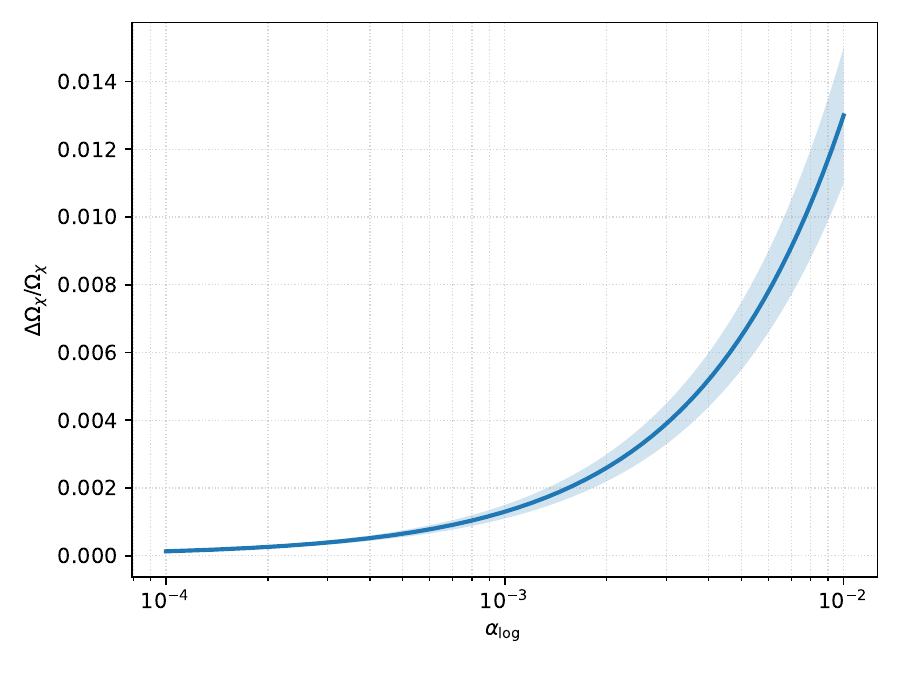}
  \caption{
  Fractional change in the relic abundance as a function of
  $\alpha_{\log}$ for the benchmark WIMP.
  The shaded band reflects the uncertainty associated with the onset
  scale $a_i$.
  }
  \label{fig:DeltaOmega_alpha}
\end{figure}

Figure~\ref{fig:DeltaOmega_alpha} shows the fractional shift
$\Delta\Omega_\chi/\Omega_\chi$ as a function of $\alpha_{\log}$.

For $\alpha_{\log}\lesssim 10^{-3}$ the effect is at the sub-percent
level, while for $\alpha_{\log}\sim 10^{-2}$ it reaches a few percent.
A precise gamma-ray determination of the relic abundance would therefore
directly constrain the CET~$\Omega$ parameter space.

%%%%%%%%%%%%%%%%%%%%%%%%%%%%%%%%%%%%%%%%%%%%%%%%%%%%%%%%%%%%%%
\section{Gamma-Ray Morphology and the Informational Field}
\texorpdfstring{$\Phi_\Omega(x)$}{PhiOmega(x)}
\label{sec:phenomena}
%%%%%%%%%%%%%%%%%%%%%%%%%%%%%%%%%%%%%%%%%%%%%%%%%%%%%%%%%%%%%%

The CET~$\Omega$ framework predicts the existence of a state-dependent
scalar quantity $\Phi_\Omega(x)$ associated with the modular–
informational sector \cite{Balfagon2026CETOmega,Haag1996}.
While its impact on the homogeneous expansion history is entirely
encoded in the parameter $\alpha_{\log}$, spatial fluctuations of
$\Phi_\Omega$ induce small but characteristic modulations of the dark
matter annihilation rate.
These modulations alter the morphology of the gamma-ray signal in a way
that is not degenerate with standard astrophysical uncertainties
\cite{Bergstrom1998,Cirelli2011,Bringmann2012}.

%%%%%%%%%%%%%%%%%%%%%%%%%%%%%%%%%%%%%%%%%%%%%%%%%%%%%%%%%%%%%%
\subsection{Evolution of the Informational Mode}
\label{subsec:phi_evolution}
%%%%%%%%%%%%%%%%%%%%%%%%%%%%%%%%%%%%%%%%%%%%%%%%%%%%%%%%%%%%%%

During the modular–informational regime ($T\sim 10$--$100$~GeV), the
field $\Phi_\Omega$ obeys the linearized equation of motion
\begin{equation}
\ddot{\Phi}_\Omega
+ 3H_\Omega\dot{\Phi}_\Omega
+ c_\psi^2\frac{k^2}{a^2}\Phi_\Omega
+ m_\psi^2\Phi_\Omega
= S_\Omega,
\label{eq:phi_eom_mod}
\end{equation}
where $m_\psi\ll H_\Omega$ is an effective mass generated by modular
dynamics, $c_\psi$ is the sound speed of informational perturbations, and
$S_\Omega$ is a stochastic source induced by renormalized modular
fluctuations \cite{Mukhanov2005,Weinberg2008}.

For super-horizon modes ($k\ll aH_\Omega$), gradient terms are negligible
and Eq.~\eqref{eq:phi_eom_mod} reduces to
\begin{equation}
\ddot{\Phi}_\Omega
+ 3H_\Omega\dot{\Phi}_\Omega
+ m_\psi^2\Phi_\Omega
\simeq S_\Omega.
\end{equation}

Since $m_\psi\ll H_\Omega$, the field undergoes a freeze-in process
analogous to curvature perturbations in standard cosmology
\cite{Mukhanov1981,Guth1981,Linde1982},
\begin{equation}
\Phi_\Omega(\bm{k},a)
\simeq
\Phi_\Omega(\bm{k},a_{\rm dec}),
\qquad (k\ll aH_\Omega),
\label{eq:phi_frozen}
\end{equation}
where $a_{\rm dec}$ denotes the scale factor at which
$\dot{\Phi}_\Omega/\Phi_\Omega\ll H_\Omega$.

Crucially, this freeze-in occurs \emph{before} WIMP freeze-out
\cite{KolbTurner1990,Steigman2012},
\begin{equation}
a_{\rm dec} < a_f,
\end{equation}
ensuring that the spatial structure of $\Phi_\Omega$ is already fixed
when dark matter departs from thermal equilibrium.

%%%%%%%%%%%%%%%%%%%%%%%%%%%%%%%%%%%%%%%%%%%%%%%%%%%%%%%%%%%%%%
\subsection{Gravitational Advection and Halo Formation}
\label{subsec:phi_advection}
%%%%%%%%%%%%%%%%%%%%%%%%%%%%%%%%%%%%%%%%%%%%%%%%%%%%%%%%%%%%%%

After freeze-in, $\Phi_\Omega$ behaves as a conserved scalar advected by
the cosmic flow \cite{Peebles1980,Bernardeau2002}.
Its late-time configuration is obtained by linear evolution from
radiation domination to the nonlinear halo regime,
\begin{equation}
\Phi_\Omega(\bm{x},a_0)
=
\int\!\frac{d^3k}{(2\pi)^3}\,
T_\Phi(k)\,
\Phi_\Omega(\bm{k},a_{\rm dec})\,
e^{i\bm{k}\cdot\bm{x}},
\label{eq:phi_transfer}
\end{equation}
where $T_\Phi(k)$ is the corresponding transfer function
\cite{EisensteinHu1998}.

The key properties of $\Phi_\Omega$ are:

\begin{enumerate}
\item \textbf{No decay:}
The informational field does not redshift away and survives to
$z=0$.

\item \textbf{Gravitational advection:}
$\Phi_\Omega$ follows the same large-scale flows that govern baryonic
and cold dark matter perturbations \cite{Peebles1980}.

\item \textbf{Finite coherence scale:}
Small-scale power is suppressed below
\[
k_\psi^{-1}\sim \frac{c_\psi}{H_{\rm dec}},
\]
but the field remains coherent on all galactic scales relevant for
annihilation signals.
\end{enumerate}

%%%%%%%%%%%%%%%%%%%%%%%%%%%%%%%%%%%%%%%%%%%%%%%%%%%%%%%%%%%%%%
\subsection{Modification of the Gamma-Ray Annihilation Rate}
\label{subsec:gamma_corr}
%%%%%%%%%%%%%%%%%%%%%%%%%%%%%%%%%%%%%%%%%%%%%%%%%%%%%%%%%%%%%%

In the standard WIMP scenario, the gamma-ray production rate per unit
volume is
\begin{equation}
\Gamma_\gamma^{(0)}(x)\propto \rho_{\rm DM}^2(x)
\end{equation}
\cite{Bergstrom1998,Cirelli2011}.

In CET~$\Omega$, the presence of $\Phi_\Omega$ induces a correction,
\begin{equation}
\Gamma_\gamma(x)
\propto
\rho_{\rm DM}^2(x)\,
\big[1+\beta_\Omega\Phi_\Omega(x)\big],
\label{eq:Gamma_corr}
\end{equation}
where $\beta_\Omega$ parametrizes the coupling between modular
fluctuations and the annihilation amplitude.

The coefficient $\beta_\Omega$ is not an independent parameter.
Both $\beta_\Omega$ and $\alpha_{\log}$ originate from insertions of the
modular Hamiltonian $K_\Omega$ in the effective action, and no additional
relevant or marginal operators couple $\Phi_\Omega$ to the annihilation
process \cite{Haag1996}.
Consequently,
\begin{equation}
\beta_\Omega = \mathcal{O}(1)\,\alpha_{\log}.
\label{eq:beta_relation}
\end{equation}

CET~$\Omega$ therefore remains an effectively \emph{one-parameter}
extension of standard cosmology \cite{Planck2018,Balfagon2026CETOmega}.

%%%%%%%%%%%%%%%%%%%%%%%%%%%%%%%%%%%%%%%%%%%%%%%%%%%%%%%%%%%%%%
\subsection{Corrected \texorpdfstring{$J$}{J}-Factor and Morphology}
%%%%%%%%%%%%%%%%%%%%%%%%%%%%%%%%%%%%%%%%%%%%%%%%%%%%%%%%%%%%%%

The line-of-sight integrated $J$-factor becomes
\begin{equation}
J_\Omega(\theta)
=
\int_{\rm l.o.s.}\! ds\,
\rho_{\rm DM}^2(s,\theta)
\big[1+\beta_\Omega\Phi_\Omega(s,\theta)\big].
\label{eq:J_factor}
\end{equation}

To linear order, the relative correction is
\begin{equation}
\frac{\Delta J}{J}
\simeq
\beta_\Omega
\langle\Phi_\Omega\rangle_{\rm halo},
\end{equation}
where the average is weighted by $\rho_{\rm DM}^2$.

For $\alpha_{\log}\sim 10^{-3}$ one finds
\[
\left|\frac{\Delta J}{J}\right|
\sim 10^{-3}\text{--}10^{-2},
\]
corresponding to sub-percent but coherent morphological distortions of
the gamma-ray signal\cite{Boddy2018Jfactor}.

%%%%%%%%%%%%%%%%%%%%%%%%%%%%%%%%%%%%%%%%%%%%%%%%%%%%%%%%%%%%%%
\subsection{Breaking Degeneracies with Other Nonstandard Cosmologies}
\label{subsec:degeneracies}
%%%%%%%%%%%%%%%%%%%%%%%%%%%%%%%%%%%%%%%%%%%%%%%%%%%%%%%%%%%%%%

Several nonstandard cosmological scenarios can modify WIMP freeze-out,
including Early Dark Energy and fast-expansion models that have been
extensively reviewed in the recent literature
\cite{Poulin2019EDE,Verde2024_ATaleOfManyH0,Salati2003FastExpansion}
,
but none reproduce the full set of CET~$\Omega$ predictions:

\begin{enumerate}
\item \textbf{Early Dark Energy:}
Affects $H(a)$ at keV temperatures, not at the GeV scale relevant for
freeze-out.

\item \textbf{Kination or fast expansion:}
Produces power-law corrections to $H(a)$ rather than a doubly logarithmic
dependence.

\item \textbf{Extra relativistic species ($\Delta N_{\rm eff}$):}
Leads to a constant fractional shift in $H$, not a scale-dependent
log--log structure.

\item \textbf{Dark-sector mediators or Sommerfeld enhancement:}
Modify $\langle\sigma v\rangle$ and spectra, but do not induce coherent
large-scale morphological changes.
\end{enumerate}

CET~$\Omega$ is uniquely characterized by the combination of a
doubly-logarithmic correction to the expansion rate, a frozen-in
informational field, and a single fundamental parameter
$\alpha_{\log}$.

%%%%%%%%%%%%%%%%%%%%%%%%%%%%%%%%%%%%%%%%%%%%%%%%%%%%%%%%%%%%%%
\subsection{Observational Prospects}
\label{subsec:observables}
%%%%%%%%%%%%%%%%%%%%%%%%%%%%%%%%%%%%%%%%%%%%%%%%%%%%%%%%%%%%%%

The morphological modulation induced by $\Phi_\Omega$ leads to:

\begin{itemize}
\item sub-percent anisotropies in the Galactic halo profile,
\item small but systematic shifts in the inferred annihilation cross
section,
\item scale-dependent deviations in the angular power spectrum of the
diffuse gamma-ray background,
\item signatures distinct from unresolved substructure boosts.
\end{itemize}

Next-generation MeV--GeV missions such as AMEGO-X, e-ASTROGAM, GECCO and
GRAMS
\cite{AMEGOX2022,eASTROGAM2018,Coogan2023GECCO,GRAMS2021},together with recent projections for precision gamma-ray morphology
studies,
are expected to reach the sensitivity required to probe

\[
\left|\Delta J/J\right|\sim 10^{-3},
\]
placing the CET~$\Omega$ morphological signal at the threshold of
detectability.

%%%%%%%%%%%%%%%%%%%%%%%%%%%%%%%%%%%%%%%%%%%%%%%%%%%%%%%%%%%%%%
\section{Conclusions}
\label{sec:conclusions}
%%%%%%%%%%%%%%%%%%%%%%%%%%%%%%%%%%%%%%%%%%%%%%%%%%%%%%%%%%%%%%

We have investigated the impact of CET~$\Omega$, an informational and
modular extension of relativistic quantum field theory and cosmology, on
the radiation-dominated era of the early Universe and on the thermal
freeze-out of weakly interacting massive particles
\cite{KolbTurner1990,Arcadi2018}.
The framework predicts a universal, state-dependent modification of the
radiation energy density in the form of a doubly logarithmic correction,
\[
\rho_\Omega(a)
=
\rho_r(a)
\left[
1+\alpha_{\log}\log\!\log\!\left(\frac{a}{a_i}\right)
\right],
\]
which arises from renormalized modular fluctuations of the spectral
triple
\cite{Connes1994,Haag1996}.

Because the correction grows extremely slowly, it leaves Big Bang
Nucleosynthesis, the Cosmic Microwave Background, and late-time
cosmological observables essentially unchanged, while becoming relevant
at the GeV temperatures characteristic of WIMP freeze-out
\cite{Pitrou2018BBN,Planck2018}.
For natural values
$10^{-4}\lesssim\alpha_{\log}\lesssim 10^{-2}$,
the modification induces percent-level shifts in the relic abundance,
fully consistent with current Planck constraints but potentially
testable through precise indirect-detection measurements
\cite{Berlin2016EarlyExpansion,Arcadi2018}.

A distinctive prediction of CET~$\Omega$ is the emergence of a frozen-in
informational scalar field $\Phi_\Omega(x)$.
This field is generated during the modular regime, freezes in before
dark matter decouples, and is subsequently advected by gravitational
collapse without decay, in close analogy with conserved cosmological
perturbations
\cite{Mukhanov2005}.
Its spatial structure induces coherent, sub-percent modulations of the
gamma-ray annihilation morphology through a derived coupling
$\beta_\Omega=\mathcal{O}(1)\alpha_{\log}$, leaving CET~$\Omega$ as an
effectively one-parameter extension of standard cosmology
\cite{Boddy2018Jfactor}.

The combined presence of:
(i) a doubly logarithmic correction to the Hubble rate,
(ii) a frozen-in informational field affecting the $J$-factor,
and (iii) the absence of additional free parameters,
renders CET~$\Omega$ distinguishable from other nonstandard cosmological
scenarios, including Early Dark Energy, kination, extra relativistic
species, or dark-sector mediator models
\cite{Poulin2019EDE,Berlin2016EarlyExpansion}.

If the reported $\sim 20$~GeV Galactic-center excess is confirmed as a
dark matter signal, it would provide a direct phenomenological probe of
$\alpha_{\log}$ and, more broadly, of the informational sector underlying
CET~$\Omega$
\cite{Daylan2016,Calore2015,Macias2018}.
More generally, the framework predicts correlated signatures across
cosmology and indirect detection that can be tested with upcoming
MeV--GeV gamma-ray missions
\cite{AMEGOX2022,eASTROGAM2018,Coogan2023GECCO,GRAMS2021}.

Finally, we note that direct-detection experiments are not expected to
be significantly affected.
The CET~$\Omega$ correction modifies the early-Universe expansion
history and the large-scale distribution of dark matter, but does not
alter local scattering amplitudes or particle-physics interactions at
late times
\cite{Arcadi2018}.

The particle-like language adopted throughout the freeze-out analysis is therefore meant as a mesoscopic effective parametrization within CET~$\Omega$.

%%%%%%%%%%%%%%%%%%%%%%%%%%%%%%%%%%%%%%%%%%%%%%%%%%%%%%%%%%%%%%
\section*{Acknowledgments}
%%%%%%%%%%%%%%%%%%%%%%%%%%%%%%%%%%%%%%%%%%%%%%%%%%%%%%%%%%%%%%

The author thanks colleagues in cosmology and astroparticle physics for
valuable discussions and feedback.
Numerical calculations and figures were produced using open-source
computational tools.
This work made use of no proprietary data and is fully reproducible.
The author dedicates this paper to his daughter Olivia, whose curiosity
and joy remain a constant source of inspiration.
%%%%%%%%%%%%%%%%%%%%%%%%%%%%%%%%%%%%%%%%%%%%%%%%%%%%%%%%%%%%%%
\appendix
%%%%%%%%%%%%%%%%%%%%%%%%%%%%%%%%%%%%%%%%%%%%%%%%%%%%%%%%%%%%%%

%%%%%%%%%%%%%%%%%%%%%%%%%%%%%%%%%%%%%%%%%%%%%%%%%%%%%%%%%%%%%%
\section{Spectral-Triple Origin of the Doubly Logarithmic Correction}
\label{app:spectral}
%%%%%%%%%%%%%%%%%%%%%%%%%%%%%%%%%%%%%%%%%%%%%%%%%%%%%%%%%%%%%%

The CET~$\Omega$ framework extends the algebraic formulation of
relativistic quantum field theory by introducing a modular–informational
sector encoded in a curved-spacetime spectral triple
\[
\left(
\mathcal A_\Omega,\,
\mathcal H_\Omega,\,
D_\Omega
\right).
\]
The modification of the radiation energy density arises from the
renormalized two-point function of fluctuations of the modular
Hamiltonian $K_\Omega$.

The correction to the energy density can be written as
\begin{equation}
\delta\rho_{\rm mod}(a)
\propto
\langle K_\Omega(a)\,K_\Omega(a)\rangle_{\rm ren}.
\label{eq:delta_rho_mod}
\end{equation}

Using the spectral representation of modular fluctuations in curved
spacetime,
\begin{equation}
\langle K_\Omega K_\Omega\rangle_{\rm ren}
=
\int d\mu(\lambda)\,
\frac{F(\lambda)}{\lambda^2+m_\psi^2(a)},
\label{eq:spectral_integral}
\end{equation}
where $\lambda$ denotes informational spectral modes and $F(\lambda)$
encodes the state dependence of the modular sector, one finds the
following generic properties:

\begin{enumerate}
\item The spectral density is approximately scale invariant,
\[
d\mu(\lambda)\sim \frac{d\lambda}{\lambda},
\]
reflecting the effective type-III behaviour of the modular algebra in
the thermodynamic limit.

\item The effective mass $m_\psi(a)$ varies only logarithmically with
the scale factor, as dictated by the modular flow.
\end{enumerate}

Evaluating Eq.~\eqref{eq:spectral_integral} yields
\begin{equation}
\delta\rho_{\rm mod}(a)
\propto
\log\!\left[
\log\!\left(\frac{a}{a_i}\right)
\right],
\end{equation}
which reproduces the universal doubly logarithmic structure of
Eq.~\eqref{eq:rhoOmega_main}.
This behaviour is robust, insensitive to ultraviolet details, and
entirely fixed by the modular scaling properties of the theory.

%%%%%%%%%%%%%%%%%%%%%%%%%%%%%%%%%%%%%%%%%%%%%%%%%%%%%%%%%%%%%%
\section{Freeze-Out Derivation in CET \texorpdfstring{$\Omega$}{Omega}}
\label{app:freeze}
%%%%%%%%%%%%%%%%%%%%%%%%%%%%%%%%%%%%%%%%%%%%%%%%%%%%%%%%%%%%%%

For completeness, we outline the derivation of the freeze-out correction
in CET~$\Omega$.
Starting from the Boltzmann equation,
\begin{equation}
\dot n_\chi + 3H_\Omega n_\chi
=
- \langle\sigma v\rangle
\left(n_\chi^2-n_{\rm eq}^2\right),
\end{equation}
and defining the yield $Y=n_\chi/s$ and $x=m_\chi/T$, one obtains
\begin{equation}
\frac{dY}{dx}
=
-\frac{s\langle\sigma v\rangle}{xH_\Omega}
\left(Y^2-Y_{\rm eq}^2\right).
\end{equation}

Freeze-out occurs when
\begin{equation}
Y(x_f)\simeq(1+c)\,Y_{\rm eq}(x_f),
\qquad c=\mathcal O(1).
\end{equation}

Expanding the Hubble rate as
\begin{equation}
H_\Omega(a)=H_r(a)\,[1+\delta(a)],
\qquad
\delta(a)=\alpha_{\log}\log\!\log(a/a_i),
\end{equation}
the shift in the freeze-out point is
\begin{equation}
\Delta x_f\simeq \frac12 x_f\,\delta(a_f),
\end{equation}
which directly leads to
\begin{equation}
\frac{\Delta\Omega_\chi}{\Omega_\chi}
\simeq
\frac12\,\alpha_{\log}
\log\!\log(a_f/a_i),
\end{equation}
in agreement with the analytic estimates presented in the main text.

%%%%%%%%%%%%%%%%%%%%%%%%%%%%%%%%%%%%%%%%%%%%%%%%%%%%%%%%%%%%%%
\section{Gamma-Ray Morphology Correction}
\label{app:morphology}
%%%%%%%%%%%%%%%%%%%%%%%%%%%%%%%%%%%%%%%%%%%%%%%%%%%%%%%%%%%%%%

The standard line-of-sight integrated $J$-factor is defined as
\begin{equation}
J(\theta)
=
\int_{\rm l.o.s.} ds\,
\rho_{\rm DM}^2(s,\theta).
\end{equation}

In CET~$\Omega$, the annihilation rate receives the correction
\begin{equation}
\Gamma_\gamma(x)
\propto
\rho_{\rm DM}^2(x)\,
\left[1+\beta_\Omega\Phi_\Omega(x)\right],
\end{equation}
leading to the modified $J$-factor
\begin{equation}
J_\Omega(\theta)
=
\int_{\rm l.o.s.} ds\,
\rho_{\rm DM}^2(s,\theta)
\left[1+\beta_\Omega\Phi_\Omega(s,\theta)\right].
\end{equation}

Writing $\Phi_\Omega=\Phi_0+\delta\Phi$ and retaining only linear terms,
one finds
\begin{equation}
\frac{\Delta J}{J}
=
\beta_\Omega
\frac{\int ds\,\rho_{\rm DM}^2\,\Phi_\Omega}
{\int ds\,\rho_{\rm DM}^2}
\equiv
\beta_\Omega\,
\langle\Phi_\Omega\rangle_J.
\end{equation}

Since $\beta_\Omega=\mathcal O(1)\alpha_{\log}$ and the primordial
normalization of $\Phi_\Omega$ yields
$\langle\Phi_\Omega\rangle_J\sim 10^{-3}$, the expected correction is
\begin{equation}
\left|\frac{\Delta J}{J}\right|
\sim 10^{-3}\text{--}10^{-2},
\end{equation}
within the reach of next-generation MeV--GeV gamma-ray observatories.

%%%%%%%%%%%%%%%%%%%%%%%%%%%%%%%%%%%%%%%%%%%%%%%%%%%%%%%%%%%%%%
% BIBLIOGRAPHY
%%%%%%%%%%%%%%%%%%%%%%%%%%%%%%%%%%%%%%%%%%%%%%%%%%%%%%%%%%%%%%

\bibliographystyle{elsarticle-num}
\bibliography{refs}

\end{document}